\documentclass[12pt,preprint]{aastex}

\begin{document}
\pagenumbering{arabic}

\title{SOME MUSINGS ON GALAXY MORPHOLOGY, GALACTIC COLORS AND THE ENVIRONMENTS OF GALAXIES}

\author{Sidney van den Bergh}
\affil{Dominion Astrophysical Observatory, Herzberg Institute of Astrophysics, National Research Council of Canada, 5071 West Saanich Road, Victoria, BC, V9E 2E7, Canada}
\email{sidney.vandenbergh@nrc.gc.ca}

\begin{abstract}

Careful inspection of large-scale photographs of Shapley-Ames
galaxies seems to show a smooth transition between the morphological
characteristics of galaxies located on the narrow red, and on the
broad blue, sequences in the galaxian color-magnitude diagram. In
other words there does not appear to be a dichotomy between blue
and red galaxies. Both the colors and the morphologies of galaxies
are found to correlate strongly with their environments. Red and
early-type Shapley-Ames galaxies are dominant in clusters, whereas
blue late-type star forming objects dominate the general field.
Interestingly the colors and morphologies of galaxies in small
groups resemble the field and differ from those in clusters. As
noted by Baade the presence of dust and star formation are very
closely correlated, except in a few galaxies that probably had
unusual evolutionary histories. Over the entire range from S0
to Sc there is no significant difference between the integrated
colors of normal and barred objects suggesting that the formation
of a bar does not significantly affect the stellar evolutionary history of
a galaxy.

\end{abstract}

\keywords {galaxies: fundamental parameters}

\section{INTRODUCTION}

In his pioneering investigation of galaxy morphology Hubble
(1926) showed that extragalactic nebulae appear to fall along a
continuous sequence that extends from his ``early'' type galaxies
of type E to ``late'' type spirals of type Sc. Various modifications
and improvements to this basic scheme were subsequently made by
Hubble (1936), de Vaucouleurs (1959), van den Bergh (1960a,b) and
Sandage (1961). This so-called Hubble sequence, and the ``tuning-fork'' diagram that takes into account the dichotomy between normal and
barred spirals, has dominated our thinking about galaxy morphology
(van den Bergh 1998, Sandage 2005) for the last three quarters of
the Twentieth Century. Very recently the availability of enormous
homogeneous surveys of galaxies has, however, suggested that the
diversity of galaxy morphologies may exhibit a previously unsuspected
dichotomy between objects that fall along blue and red sequences in
the luminosity versus color diagram. This effect is beautifully
shown in Consolice (2006), and even more clearly in  Wang
et al. (2007). Recent detailed bibliographies of work on this
subject are provided  by Cattaneo et al. (2006) and Renzini (2006).
Furthermore Faber et al. (2006) show that this apparent
dichotomy between the blue and red sequences of galaxies extends
back in time to a redshift of at least z = 1.0.  Arnouts et al. (2007) find evidence for a major buildup of the red sequence for $1 < z < 2$. Finally Lab$\acute{e}$ et al. (2007) observe that the red sequence apears to be absent at $z \sim$ 3.

The dependence of
the relative populations along the blue and red sequences on
galaxy luminosity is discussed in detail by Baldry et al. (2004).
Baldry et al. (2006) also provide interesting information on
the dependence of the relative strengths of the blue and red
galaxy sequences on environmental density. To explain the observed
bi-modality in the color-luminosity diagram it has been suggested
that, after a critical epoch at z $\sim$ 3, only those dark matter halos
below a critical shock-heating mass of $\sim 10^{12}$ $M_\odot$ enjoyed
inflow of cold gas that could form stars, whereas cooling and star
formation were shut down abruptly above this mass. According to
Cattaneo et al., and references cited therein, galaxies arrive at
the bright end of the red sequence by dissipationless (``dry'')
mergers, or alternatively via ``wet'' mergers among the most luminous
galaxies of the blue sequence. It is one of the purposes of the
present investigation to search for possible systematic morphological
differences between galaxies on these blue and red sequences. In
particular it is attempted to answer the question asked by Ball
et al. 2006): Does the morphology of galaxies reveal anything that
colors do not?

\section{THE DATA}

In his recent investigation Conselice (2006) used a data base
consisting of the physical characteristics of a large (22 121
galaxies), but quite inhomogeneous, database drawn from the RC3
catalog of de Vaucouleurs et al. (1991). The present investigation
is based on the much smaller, but far more homogeneous, sample
provided by the 1246 galaxies in {\it The Revised Shapley-Ames Catalog of
Bright Galaxies} (Sandage \& Tammann 1981). This catalog contains the
largest and most uniform existing collection of high quality galaxy
classifications. All of these classifications are based on inspection
of plates obtained with large reflecting telescopes. Furthermore
beautiful large-scale photographs of each of these galaxies have been
published in the monumental {\it Carnegie Atlas of Galaxies} (Sandage \&
Bedke 1994). Table 1 lists those  galaxies in the Revised Shapley -
Ames catalog for which integrated $(U - B)_{o}$ colors are also given in
the {\it Third Reference Catalogue of Bright Galaxies} of de Vaucouleurs et al. (1991). Such
$(U-B)_{o}$ colors should provide a sensitive measure of the current rate
of star formation. These colors were derived by applying statistical
corrections for both Galactic foreground reddening and internal dust
reddening. It should be noted that these corrections for reddening
are particularly uncertain for galaxies at low Galactic latitudes.
To remind the reader of this fact all U-B colors of galaxies with
$\mid b \mid < 25.0^{o}$ in Table 1 are followed by a colon (:). By the same
token the statistical corrections to de Vaucouleurs et al. ``face-on''
colors are, of course, particularly large and uncertain for dusty
almost edge-on spirals such as M31. Also listed for all of the
galaxies in Table 1 is a somewhat simplified version of the
morphological classifications by Sandage \& Tammann, as well as the
the absolute magnitudes $M^{o,i}_{B_T}$ (which will subsequently, for
the sake of simplicity, be referred to as $M_{B}$). Except for Local
Group members, the luminosities adopted by Sandage \& Tammann were
reduced by 0.73 mag so that they are now based on a Hubble
parameter of 70 km $s^{-1}$ Mp$c^{-1}$, rather than on the value of
50 km $s^{-1}$ Mp$c^{-1}$, that was favored by Sandage \& Tammann (1981).
Following Conselice (2006) the galaxies in the red sequence and
those in the blue sequence  were separated by an empirical
dividing line defined by the relation

 \hspace*{4cm}$M_{B}$ = -14.98 - 31.8 $(U-B)_{o}$.\hspace*{4cm}  (1)
                             
In Table 1 all galaxies that fall to the blue of this relation
are designated ``B'', whereas those that fall to the red of it have
been designated ``R''. Following Conselice (2006) a small number of
dwarf galaxies fainter than $M_{B}$ = -18.00 were excluded from the
sample and are not listed in Table 1.

The total number of galaxies contained in the final sample
listed in Table 1 is, per chance, exactly 800. For galaxies North of
$\delta$ = $-27^{o}$, the table also gives information on the environment
of each galaxy based on inspection of their images on the prints of
the Palomar Sky Survey. Objects that appeared to be accompanied by
fewer than three non-dwarf companions were provisionally assigned the
field [F], those that appeared to have 3 to 6 non-dwarf companions
were tentatively allocated to groups [G], and those that seemed to
have more than six non-dwarf companions were assumed to be cluster
[C] members. These designations were based only on visual inspection
of Sky Survey prints and not on radial velocity data. This approach
was adopted because the large random motions superimposed on the
smooth Hubble flow can, in some cases, result in misleading distance
estimates for the relatively nearby galaxies in the Shapely-Ames
Catalog. Clearly the environmental information given in Table 1 should
only be regarded as being indicative in nature. Nevertheless it is
encouraging [see Section 6] that a very strong correlation is found
between the environmental classes ``F'', ``G'' and ``C'' and the morphological
types and colors of the galaxies located in them. Since the Local Group
and the Virgo cluster extend over many Palomar Sky Survey prints,
memberships in these two features were taken from the catalog of
van den Bergh (1960c). Finally Table 1 gives visual estimates of the
amount of dust present in, and the intensity of star formation in,
all of the galaxies contained in Volume 1 (types E - Sb) of {\it The
Carnegie Atlas of Galaxies} (Sandage \& Bedke 1994).

\section{THE COLOR-LUMINOSITY DIAGRAM}

As Choi et al.  (2007) have recently reminded us abolute magnitude and morphology are the most important parameters characterizing the physical properties of galaxies.  Although the present investigation contains two orders of magnitude fewer data points than their work it provides much more accurate morphological information since the galaxies were sorted into a dozen or so morphological classes, compared to the two broad classes [E/S0 and S/Ir] used by Choi et al.  These two investigations therefore provide complimentary types of information.

In van den Bergh (2007) it was pointed out that there appears
to be a contradiction between the seeming continuity of the Hubble
sequence E - Sa - Sb - Sc - Ir and the apparent bimodality that
is exhibited by the distribution of galaxies in the color-luminosity
diagram. It is one of the purposes of the present paper to see if
the uniform and high quality of the morphological classifications
of galaxies in the Shapley-Ames catalog can throw any light on this
apparent discrepancy. Figure 1 shows a color magnitude diagram
for all of the 800 Shapley-Ames galaxies for which de Vaucouleurs
et al. (1991) give intrinsic $(U-B)_{o}$ colors, and for which Sandage
\& Tammann (1981) list luminosities and accurate morphological
classifications. The figure clearly shows that the intrinsic color-
magnitude diagram of nearby galaxies is bimodal. However, this
bimodality appears less clear-cut than that exhibited by similar data
for all RC3 galaxies with $(U-B)_{o}$ colors that are plotted in Figure 7
of Conselice (2006) and for that seen in the $M_{r}$ versus g-r diagram
plotted in Figure 2 of Wang et al. (2007). A color-magnitude diagram
for the elliptical galaxies in Table 1 is shown in Figure 2. Such E galaxies are
seen to scatter about a line that is parallel to, but $\sim$0.30 mag
redder, than Eqn. (1) which separates red and blue galaxies. A
few objects that scatter to the left (blue) of the mean relation
for ellipticals are, presumably, galaxies that still harbor a
low level of star formation. The bluest object in the Figure 2
is the starburst galaxy NGC 1275 (Perseus A). Figure 3 shows
a color-magnitude diagram for all Shapley-Ames S0 + SB0 galaxies.
The majority of these objects are seen to fall along the same
sequence as do the ellipticals in Fig. 2 but there is an
increased scatter towards bluer colors. The bluest object in the
figure is NGC 3616. Perhaps surprisingly, no obvious systematic
differences are seen between the locations of $S0_{1}$, $S0_{2}$ and $S0_{3}$
galaxies in the $M_{B}$ versus $(U-B)_{o}$ diagram. Figure 4 shows that
the majority of Sa + SBa galaxies are red, but that they exhibit a
considerable scatter in color. In particular there is no longer
any obvious concentration of objects associated with the red
sequence that is outlined (or dominated by) elliptical galaxies.
Figure 5 shows that Sb and SBb galaxies exhibit a large range
in colors, with the majority of these objects falling to the blue
of Eqn. (1). Finally Figure 6 shows that the overwhelming majority
of Sc and SBc galaxies scatter to the left (blue) of Eqn. (1). It
is interesting to note that Figures 3, 4, 5 and 6 exhibit no
systematic difference between the distribution of S0 and SB0, Sa
and SBa, Sb and SBb or Sc and SBc galaxies. Plots of Sab + SBab and
of Sbc + SBbc galaxies show that same conclusion also holds for those
objects. This observation suggests that the presence, or absence, of
a bar does not strongly affect the evolutionary history of star formation in a galaxy.
I am indebted to Lia Athanassoula for pointing out to me that the
reason for this may be that the presence of a bar merely rearranges
the material in the disk of a galaxy without profoundly affecting
its evolution. It is noteworthy that spiral galaxies within each of
the individual morphological types from Sab to Sc exhibit a
large spread in their $(U-B)_{o}$ colors. Because the morphological
classifications by Sandage \& Tammann are of such high quality this
scatter can not be attributed to errors in the morphological
classifications. It follows that there exists a significant range
in the rate of star formation within each of the morphological
classes Sab, Sb, Sbc and Sc. In other words galaxy morphology
and galaxy color appear to provide complementary information on
the evolutionary status of a galaxy. The relation between
morphology and color may be affected by both differences in the
time of onset of gas infall (Kampakoglou \& Silk 2007) and on the
more recent rate of gas accumulation.  Park et al. (2007) have recently reminded us the fact [see, for example, Figure 1 of van den Bergh (1998)] that very late-type galaxies are systematically less luminous than are galaxies of intermediate and early Hubble types.  Since early-type galaxies mainly occur in clusters, and very late-type galaxies mostly in the field, there is a systematic difference between the luminosity functions of field and cluster regions.  Figures 7, 8 and 9 also show (as has also been emphasized recently by Park et al.) that the brightest very red galaxies in clusters are more luminous that are the brightest very red field galaxies.  Finally Table 2 shows the distribution of the various sub-types of S0 galaxies in differing environments.  The table shows that $S0_{1}$ galaxies are most frequent in the cluster environment.  However, with $\chi^{2}$ = 8.9 for 4 degrees of freedom, the distribution of S0 sub-types does not appear to depend significantly on environment.

\section{COLOR AND MORPHOLOGY}

 Broadly speaking the transition between galaxies on the B
and on the R sequence in the color-luminosity diagram of galaxies
occurs between Hubble types Sab and Sb. To study this transition in
detail all of the images of galaxies listed in Table 1, that are
shown in Volume 1 (which covers morphological types E, S0, Sa, Sab
and Sb) of {\it The Carnegie Atlas of Galaxies} were inspected. These
images show that the strength of the central bulge decreases
gradually and systematically along the sequence E-Sa-Sab-Sb, while at
the same time the strength of the star forming disk population
increases gradually. Careful inspection of all of these images shows
no indication of a discontinuity in the morphological characteristics
of galaxies as one goes from objects on the R sequence to those that
fall along the B sequence. Generally speaking objects that fall to
the left of Eqn (1) are of types Sb, Sbc and Sc, whereas most of
those to the right of the line defined by Eqn. (1) are of types
E, S0, Sa and Sab. Obvious exceptions, such ase the very red
Sb galaxy NGC 3169, are seen to be exceptionally dusty. On
the other hand the rather blue Sab galaxy NGC 6887 may be a
misclassified Sb. Finally some Sb galaxies such as NGC 2841 and
NGC 7217, which have almost circular spiral arms, are found to
lie on the R sequence. In conclusion inspection of all of the available
high quality images leaves one with the impression that there is a
gradual transition in morphology along the Hubble sequence, rather
than a clear-cut dichotomy between galaxies on the B and R sequences.
The most likely explanation for the apparent dichotomy exhibited by
the color data may be that most cluster galaxies have low star
formation rates and $(U-B)_{o}$ $\sim$ +0.5, whereas the majority of group and
field galaxies exhibit a wide range in colors with a maximum at
$(U-B)_{o}$ $\sim$ -0.1.

The question ``Does morphology tell us more than galaxy colors?''
does not have a simple answer. For most galaxies the ($U-B)_{o}$ color
correlates closely with morphological type. A large deviation from
the normal relation between color and luminosity usually indicates
that a galaxy is peculiar is some way, i.e the A + K  galaxy
NGC 1275. By the same token a deviation from the normal relation
between the tilt of spiral arms, and the strength of the central
bulge, often points to a galaxy that has probably had an unusual
evolutionary history.

\section{DUST, MORPHOLOGY AND STAR FORMATION}

One of Walter Baade's favorite sayings was ``No dust, no
Population I''. Inspection of the full sample of galaxy images shown
in {\it The Carnegie Atlas of Galaxies} beautifully confirms this strong
correlation between the presence of dust and star forming activity.
Inspection of the images of all those galaxies in Table 1, that are
shown in Volume 1 (E-S0-Sa-Sb) of the {\it Carnegie Atlas of Galaxies}
(Sandage \& Bedke 1994), allows one to classify the visibility of
dust on a scale from D = 0 (dust free), D = 1 (trace of dust), D = 2
(dusty), to D = 3 (very dusty). Similarly the apparent intensity
of star formation can be graded on the scale S = 0 (no star formation),
S = 1 (trace of star formation), S = 2 (active star formation) and
S = 3 (very active star formation). Obvious caveats are that dust
and star formation may, in some images, be invisible because it
occurs in overexposed regions. Furthermore the visibility of dust
will sometimes depend on galaxy orientation (e.g. NGC 7814). Also
the evidence for star formation may be more obvious in nearby than
in distant galaxies. Nevertheless the data in Table 3 show a very
strong correlation between the indices that describe the dust and
star formation. It is, however, instructive to note some striking
exceptions to ``Baade's rule''. Metal-poor galaxies, such as the Small
Magellanic Cloud, exhibit strong star formation, yet show only dim
traces of dust absorption. In the present data set this problem
is minimized by the fact that dwarf galaxies with $M_{B}$ $>$ -18.00
(which are usually metal-poor, and hence deficient in dust) have
been excluded from the data set contained in Table 1. In other
cases such as NGC 2146 and NGC 4438 (tidal?) warping appears to
have made the dust lanes particularly prominent because spiral arms
have been lifted above most of the sources of stellar radiation.
There are also a few rather mysterious objects, such as NGC 4826,
that exhibit enormous dense dust clouds but few very bright stars.
Perhaps such galaxies resemble M82 which appears to lack bright
supergiant stars even though it contains a great deal of absorbing
dust. Possibly such galaxies had a burst of star formation that ended
suddenly a few tens of millions of years ago. Sandage \& Tammann
(1981) call some objects of this type ``amorphous'' galaxies. $S0_{3}$
galaxies are dusty, but usually without obvious evidence for
star formation. In some cases the dust distribution seems to be
chaotic (e.g. NGC 4753). In others, like NGC 2907, the dust lies
in the fundamental plane.  The color magnitude diagram for galaxies with no dust (D = 0) is shown in Figure 10, and that for galaxies with a trace of dust (D = 1) in Figure 11.  Inter-comparison of these figues suggests that many of the galaxies that exhibit a trace of dust absorption have star formation, and therefore exhibit bluer (U-B) colors, than do dustless galaxies.  Perhaps surprisingly, the color-magnitude diagrams for galaxies with strong (D = 2) or very strong (D = 3) visual dust absorption have color-magnitude diagrams that differ but little from that for galaxies that exhibit only a trace (D = 1) of dust absorption.  Possibly this observed effect is due to internal reddening in very dusty galaxies, canceling out the effects of young blue stars on the integrated colors of galaxies.  

Inspection of Figure 12 shows that those galaxies in Volume 1 of {\it The Carnegie Atlas of Galaxies}, in which no star formation is visible (S = 0), fall along a sequence that lies 0.3 mag to the red of the dividing line given by Eqn. (1).  On the other hand Figure 11 shows a much broader integrated color distribution for galaxies with S = 1 that exhibit some evidence for star formation. 

\section{GALAXY MORPHOLOGY AND ENVIRONMENTAL DENSITY}

It was first pointed out by Hubble (1936, p.81) that the nature
of galaxian populations depends on the density of their environment,
with early type galaxies dominating in dense regions of the Universe.
The nature of this relation was explored in detail by Dressler
(1980) and has most recently been discussed by Hogg et al. (2004),
who showed that bulge dominated galaxies which have large S\'{e}rsic
indices, mainly occur in dense regions. In the present investigation
all Shapley-Ames galaxies with  $\delta > -27^{o}$ were to assigned to
clusters (C), groups (G) and the general field (F) by inspection of the
prints of the Palomar Sky Survey, using the criteria described in
Section 2. As Whiting et al. (2007) have recently pointed out ``[T]he
idea of a visual survey on photographic material appears almost
quixotic''. Nevertheless this appears to be the most efficient way of
assessing the environmental characteristics of nearby galaxies, such as
those contained in the Shapley-Ames catalog.  It should, of course, be emphasized that the division of galaxies into F, G and C regions is an artificial one since, as Park et al. (2007) have emphasized recently, the properties of galaxies appear to vary smoothly as a function of environmental density.   The results of the
present survey are shown in Figures 7, 8 and 9, These figures
show that, in the $M_{B}$ versus $(U-B)_{o}$ diagram, cluster galaxies
overwhelmingly occur to the right (red) of the line defined by Eqn. (1),
whereas field galaxies mostly scatter to the left (blue) of this
relation.  From their study Park et al. (2007) find that high-density regions contain very bright galaxies, whereas low-density regions do not.  Inspection of Figures 7, 8 nad 9 appears to show a similar trend for the reddest galaxies having $(U-B)_{o} > +0.50$.  However, a Kolmogorov-Smirnov test shows that this effect does not reach a respectable level of statistical significance in the present dataset.  Furthermore, the distribution of $S0_{1}$, $S0_{2}$ and $S0_{3}$ galaxies is not found to differ significantly between cluster and field regions.

Table 4 shows the distribution of morphological types of the
galaxies listed in Table 1 as a function of environment. This table
clearly shows that early type galaxies are most frequent in clusters,
and that objects of late type predominate in the field. A Kolmogorov-
Smirnov test shows that there is $<$ 0.01\% probability that early-type
and late-type galaxies have the same relative frequency distribution
in clusters and in the field. Finally Table 5 shows that the galaxies in
clusters are mainly red and that those in the field are predominantly
blue. A K-S test shows that there is $<$0.01\% probability that the
distribution of intrinsic colors in clusters and in the field have
been drawn from the same parent distribution of intrinsic colors.
Similarly K-S tests of the data listed in Tables 4 and 5 show
probabilities of 0.6\% and 0.2\% respectively for the similarity of
morphological types and intrinsic colors of galaxies in clusters
and in groups. In other words there is a real, and highly significant,
difference between the galaxian content of groups and clusters. On
the other hand K-S tests of the data in Tables 4 and 5 show that
the distributions of morphological types, and of the colors of galaxies
in groups and in the field, do not differ at a respectable level of
statistical significance. This suggests that groups of galaxies (such
as the Local Group) have galaxian populations that resemble those
of isolated field galaxies. In other words groups appear to
be density enhancements within the field. It is also interesting to
note that the study of the Hubble morphological types of galaxies,
and investigations of the intrinsic colors of galaxies yield almost
identical results. In other words it is not possible to say if
galaxy color provides a more significant way of characterizing a
galaxy than does its morphological type.

\section{CONCLUSIONS}

Large-scale photographs of all non-dwarf early-type galaxies in
The Shapley-Ames Catalog, for which $(U-B)_{o}$ colors are available, have
been carefully inspected to search for possible morphological
discontinuities between objects that lie on the relatively narrow red
sequence and the broader blue sequence in the galaxian color-luminosity
diagram. Furthermore all of these objects were graded on the basis of
their dust content and apparent rate of star formation. Finally the
environments of all Shapley-Ames galaxies with  $\delta > -27^{o}$ were inspected on the prints of the Palomar Sky Survey to see if they appeared to
belong  to clusters, to groups, or to the general field.

\subsection{Red and blue sequences.}

Careful inspection of the morphologies of all of the Shapley-Ames
galaxies listed in Table 1 strongly suggests that: (1) the strengths
of central bulges decrease smoothly towards later types and bluer
colors while, (2) the strength of the disk component increases gradually
towards later Hubble types and bluer colors. In other words there
appears to be no obvious dichotomy between the morphologies of
galaxies that are situated on the broad blue and on the narrow red
sequences in the galaxian color-magnitude diagram. Large deviations from the
average relationships between the colors and the morphological types of
galaxies are often found to be indicative of an unusual evolutionary
histories.

\subsection {Galaxy colors and environments.}

Both the colors and morphological types of galaxies are seen
to correlate strongly with their environments, with galaxies on
the red sequence being dominant in clusters and objects on the
blue sequence predominating in general field. Interestingly galaxies
located in groups (such as the Local Group) are found to have
colors and morphologies resembling field galaxies. In other
words groups appear to be condensations within the field, rather than
close relatives of clusters. Perhaps this is what Hubble (1936,p.126)
meant when he wrote that the Local Group was ``[A] typical, small group
of nebulae which is isolated in the general field.''

\subsection{Dust and star formation.}

Inspection of the information for 404 early-type galaxies having
estimates of both dustiness and intensity of star formation strongly
confirms Walter Baade's conclusion about the close correlation between
the presence of dust and star formation. A few dusty galaxies, with
no obvious star formation, are probably objects that have had an
unusual evolutionary history.

\subsection{Normal and barred galaxies.}

No systematic color differences were found between S0 and SB0,
Sa and SBa, Sb and SBb and Sc and SBc galaxies. This shows that
the presence or absence of a bar does not strongly affect the stellar
evolutionary history of a galaxy. The reason for this may be
that a bar merely rearranges material in the disk without
profoundly affecting its evolution.

\subsection{Properties of S0 galaxies.}

Surprisingly the relative frequency of
$S0_{1}$, $S0_{2}$ and $S0_{3}$ galaxies does not appear to differ significantly
between cluster, group and field regions.

\subsection{Dust and galaxy colors}

The presence/absense of obvious signs of star formation and the presence/absence of dust affect the $(U-B)_{0}$ colors of galaxies in the expected way.

I am deeply indebted to Michael Peddle for plotting the figures in
this paper and to Brenda Parrish for technical assistance with the
manuscript. I am indebted to Harold Corwin for a very helpful and
detailed discussion on some identification problems in the catalog of
Sandage \& Tammann. Thanks are also due to Lia Athanassoula for her
comments on the nature of barred spirals, and to John Kormendy and Niv Drory for discussions of the formation of galactic bulges.  Finally I am indebted to the referee, Michael Vogeley, for a number of helpful suggestions.

\begin{deluxetable}{lclllccc}
\tabletypesize{\small}
\tablewidth{0pt}        
\tablecaption{Catalog of Shapley-Ames galaxies with $M_{B} < -18.0$ for which de Vaucoulers et al. (1991) list integrated $(U-B)_{o}$ colors.} 

\tablehead{\colhead{Name} & \colhead{Sequence} &  \colhead{Type} &  \colhead{$M_{B}$}  &  \colhead{$(U-B)_{o}$}  &  \colhead{Environment}  & \colhead{D}   &  \colhead{S}}

\startdata

N0016 &   R &     SB0 1        & -21.54 & +0.37  &     C   &     0 &  0   \nl
N0024 &   B &     Sc II-III    & -18.39 & -0.17  &     F   &	   &	  \\
N0055 &   B &     Sc           & -19.42 &  0.00  &    ...  &	   &	  \\
N0095 &   B &     Sc II        & -21.43 & -0.05  &     F   &	   &	  \\
N0128 &   R &     S0 2 p       & -21.38 & +0.53  &     G   &     0 &  0	  \\
N0134 &   B &     Sbc II-III:  & -21.55 & +0.11  &    ...  &	   &	  \\
N0151 &   B &     SBbc II      & -21.97 & +0.04  &     F   &	   &	  \\
N0148 &   R &     S0 2         & -19.15 & +0.46  &    ...  &     0 &  0	  \\
N0150 &   B &     Sbc II       & -20.58 & -0.05  &    ...  &	   &	  \\
\enddata
\tablenotetext{*}{I am indebted to Dr. Harold Corwin for pointing out that object ''New 5'' in the catalog of Sandage \& Tammann is probably identical to IC 4946.  The morphological type assigned to NGC 2798 might  be questioned NGC 3415 is classified as E5 in Sandage \& Tammann, but as Sb in Sandage \& Bedke.}
\end{deluxetable}

\clearpage

\begin{deluxetable}{lccc}
\tablewidth{0pt}
\tablecaption{Relation between S0+SB0 subtype and environment}

\tablehead{\colhead{Environment}  &   \colhead{$S0_{1}$}  & 
\colhead{$S0_{1/2} + S0_{2}$}  & \colhead{$S0_{2/3} + S0_{3}$}}
\startdata

Field        &    9   &   9   &  3\\

Group        &    7   &   2   &  4\\

Cluster      &    28  &   6  &   5\\
\enddata
\end{deluxetable}

\begin{deluxetable}{lllll}
\tablewidth{0pt}        
\tablecaption{ Relation between presence of dust and the apparent intensity of star formation} 

\tablehead{\colhead{} & \colhead{S = 0}  & \colhead{S = 1}  & \colhead{S = 2}   & \colhead{S = 3}}

\startdata

D=0 &  183  &  37   &  4   &  0\\

D=1 &  10   &  31   &   37  &   1\\

D=2  &   11 &   35  &  35   &  4\\

D=3  &    0  &   7  &   7  &   2\\

\enddata
\end{deluxetable}

\begin{deluxetable}{llll}
\tablewidth{0pt}        
\tablecaption{Frequency distribution of Hubble types in Table 1} 

\tablehead{\colhead{Type} & \colhead{Cluster} & \colhead{Group} & \colhead{Field}}

\startdata

E         &     43    &    12   &    12\\

S0 + SB0  &     34    &    9    &  19\\

Sa + SBa  &  19    &    7  &    26\\

Sab + SBab  &  10   &      1    &  11\\

Sb + SBb &    19   &     13   &   33\\

Sbc + SBbc    &  10    &     7  &    27\\

Sc + SBc  &   33    &   25   &   80\\

\enddata
\end{deluxetable}

\begin{deluxetable}{llll}
\tablewidth{0pt}        
\tablecaption{ Frequency distribution of intrinsic galaxy
colors in Table 1 in bins of width 0.10 mag} 

\tablehead{\colhead{$(U-B)_{o}$} & \colhead{Cluster} & \colhead{Group} & \colhead{Field}}

\startdata
-0.45  &    2   &      4  &     6\\
-0.35  &    4   &      1  &     4\\
-0.25  &    7  &       3  &    21\\
-0.15  &   11  &      10  &    39\\
-0.05  &   20  &      10  &    43\\
+0.05  &   10  &      16  &    34\\
+0.15 &    20  &       6  &     25\\
+0.25  &   28  &      10  &    28\\
+0.35  &   29  &       8  &    21\\
+0.45  &   50  &      14  &    28\\
+0.55  &   36  &      12  &    13\\
+0.65  &    4  &       1  &     2\\

\enddata
\end{deluxetable}

\clearpage

\begin{figure}
\plotone{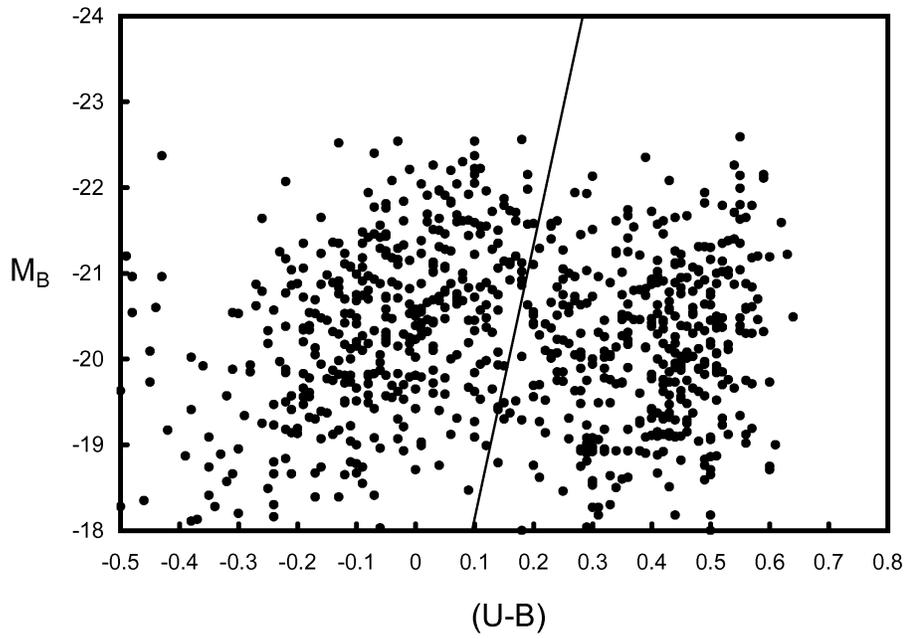}
\caption{Color-magnitude diagram for all Shapley-Ames galaxies for which $M_{B}$ and $(U-B)_{o}$ are available from de Vaucouleurs et al.(1991) and Sandage \& Tamman (1981).  The figure shows that the distribution of galaxy colors is bimodal.}
\end{figure}

\begin{figure}
\plotone{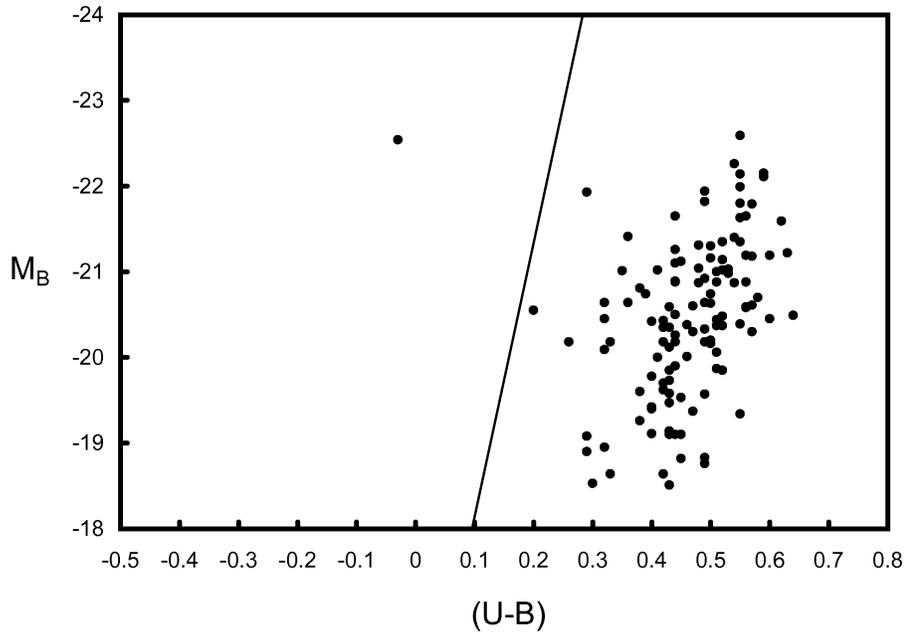}
\caption{Color-magnitude diagram for elliptical galaxes. The deviant blue point represents NGC 1275.}
\end{figure}

\begin{figure}
\plotone{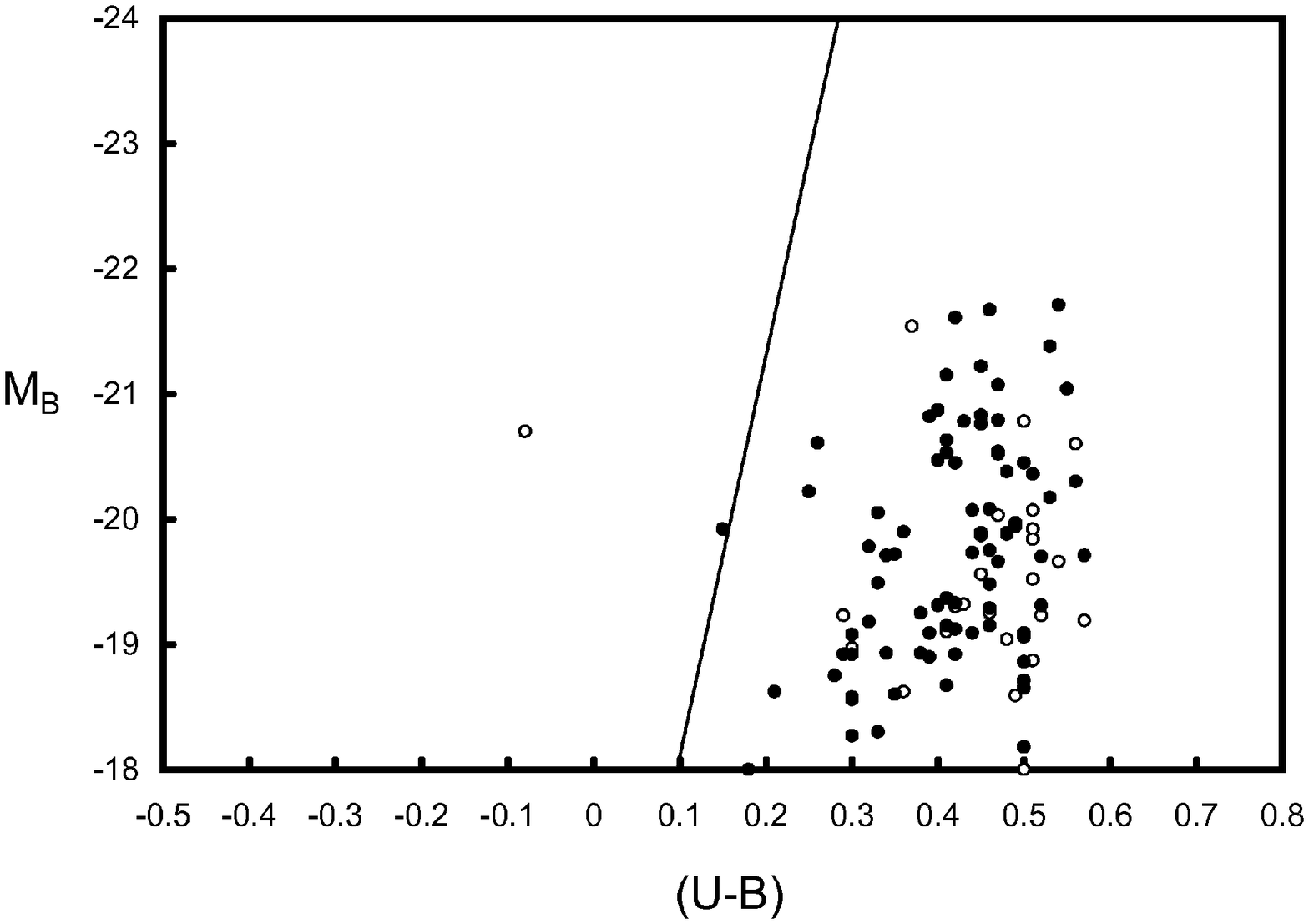}
\caption{Color-magnitude diagram for S0 galaxies (dots) and SB0 galaxies (circles).}
\end{figure}

\begin{figure}
\plotone{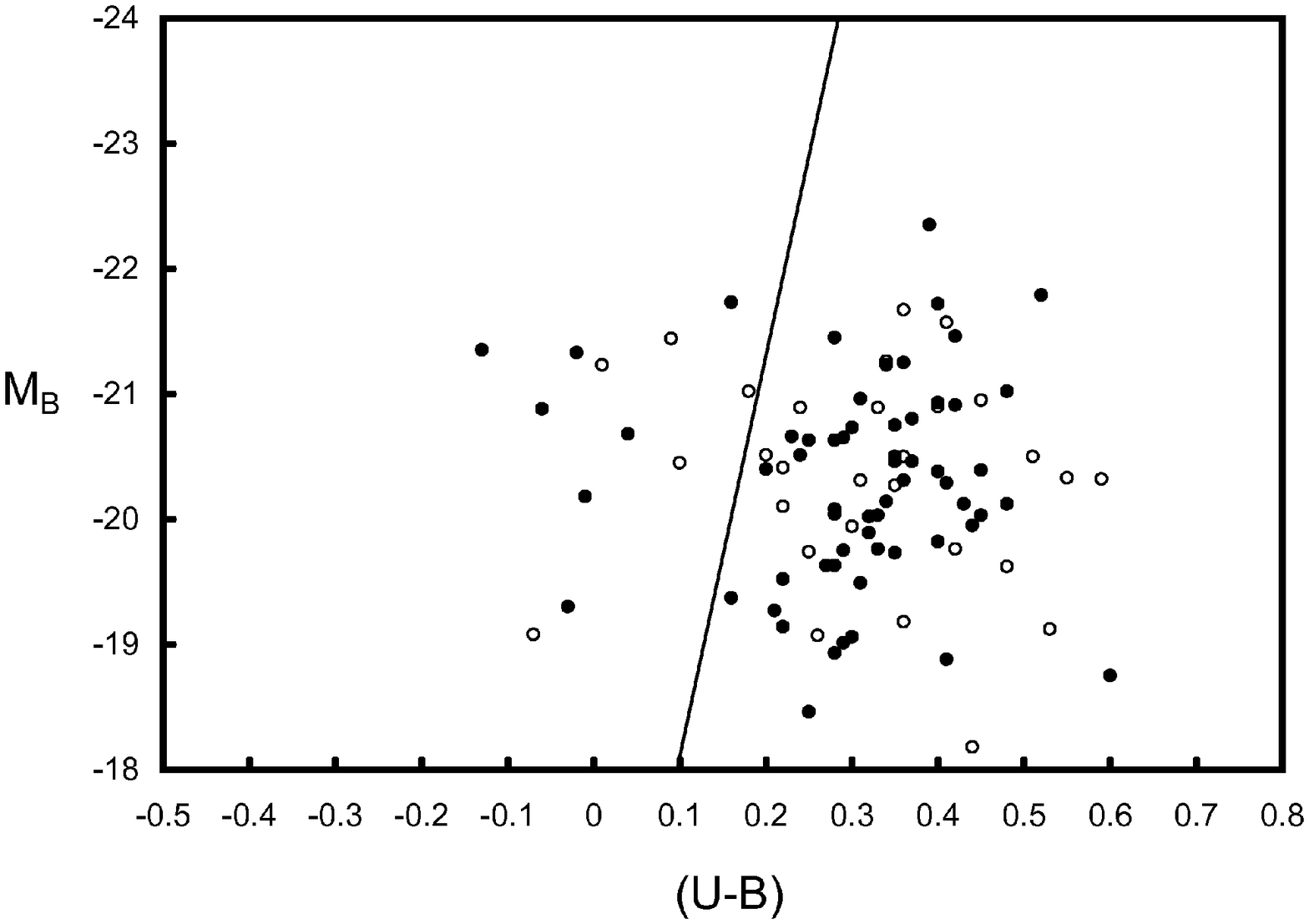}
\caption{Color-magnitude diagram for Sa galaxies (dots) and SBa galaxies (circles).}
\end{figure}

\begin{figure}
\plotone{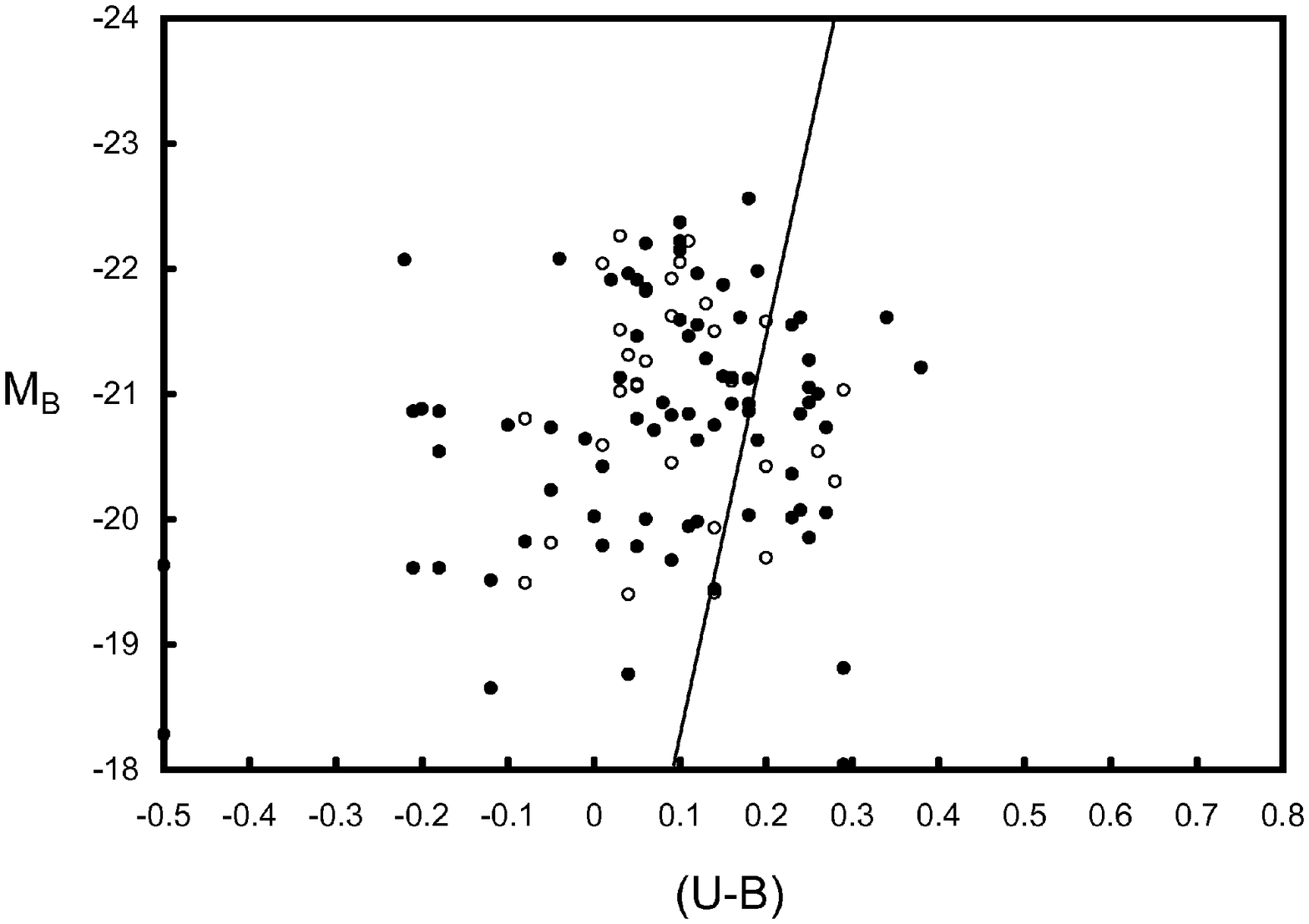}
\caption{Color-magnitude diagram for Sb galaxies (dots)and SBb galaxies (circles).}
\end{figure}

\begin{figure}
\plotone{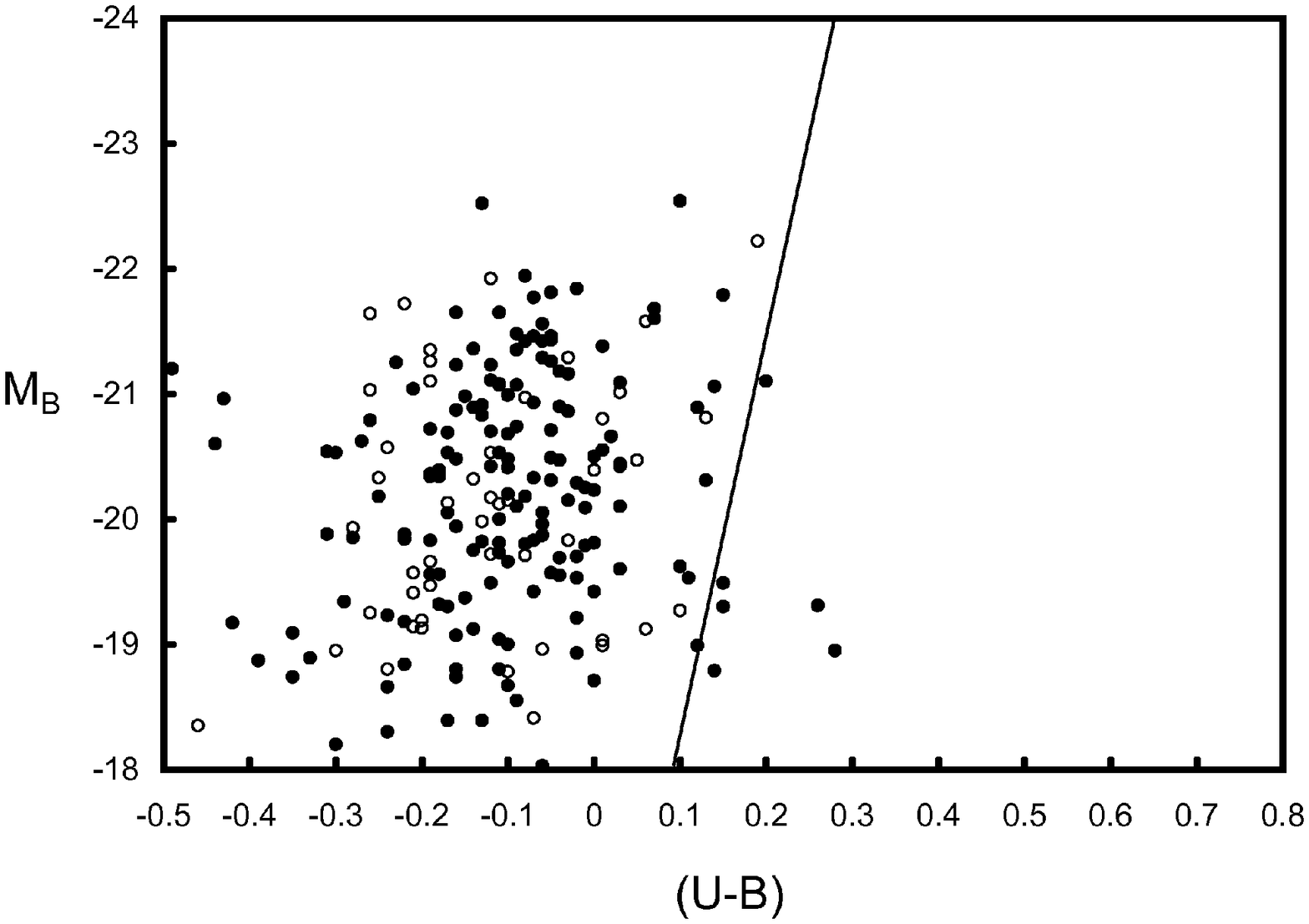}
\caption{Color-magnitude diagram for Sc galaxies (dots) and SBc galaxies (circles).}
\end{figure}

\begin{figure}
\plotone{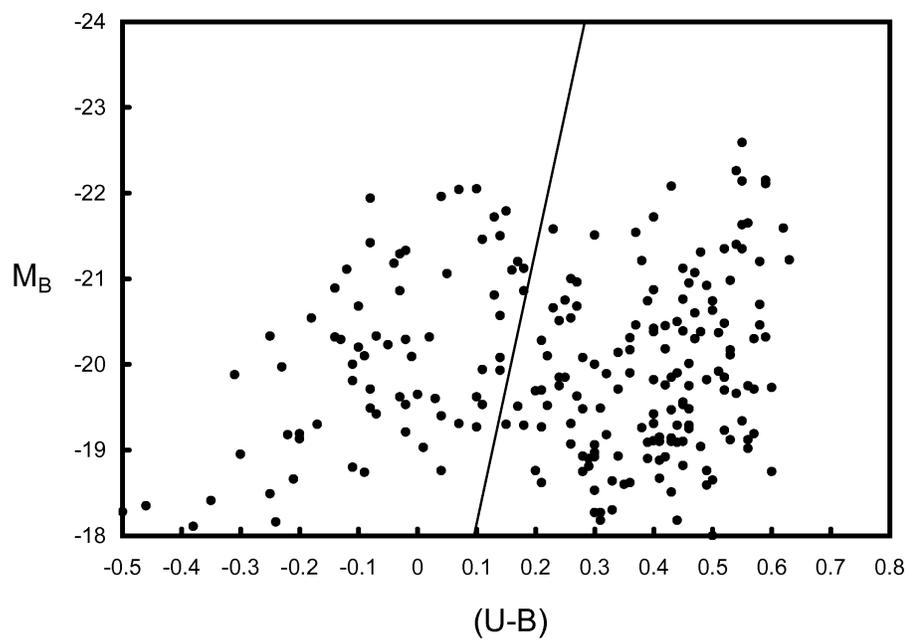}
\caption{Color-magnitude diagram for cluster (C) galaxies in the Shapley-Ames catalog. These objects are seen to mostly be located to the red of Eqn. (1).}
\end{figure}

\begin{figure}
\plotone{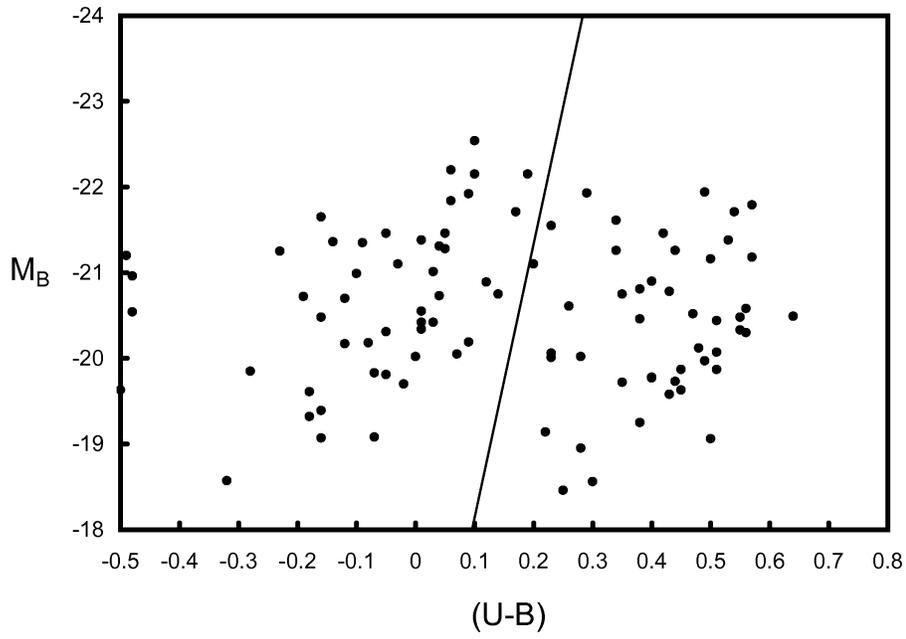}
\caption{Color-magnitude diagram for group (G) galxies in the Shapley-Ames catalog.}
\end{figure}

\begin{figure}
\plotone{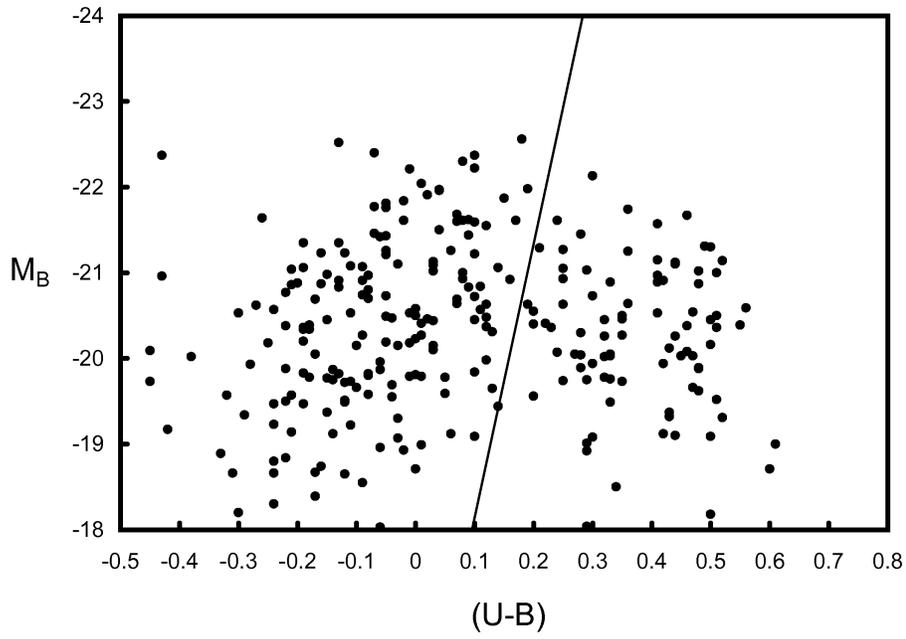}
\caption{Color-magnitude diagram for field (F) galaxies. These objects are seen to be widely scatteded in the color-luminosity diagram, but mainly lie to the left (blue) of Eqn.(1).}
\end{figure}

\begin{figure}
\plotone{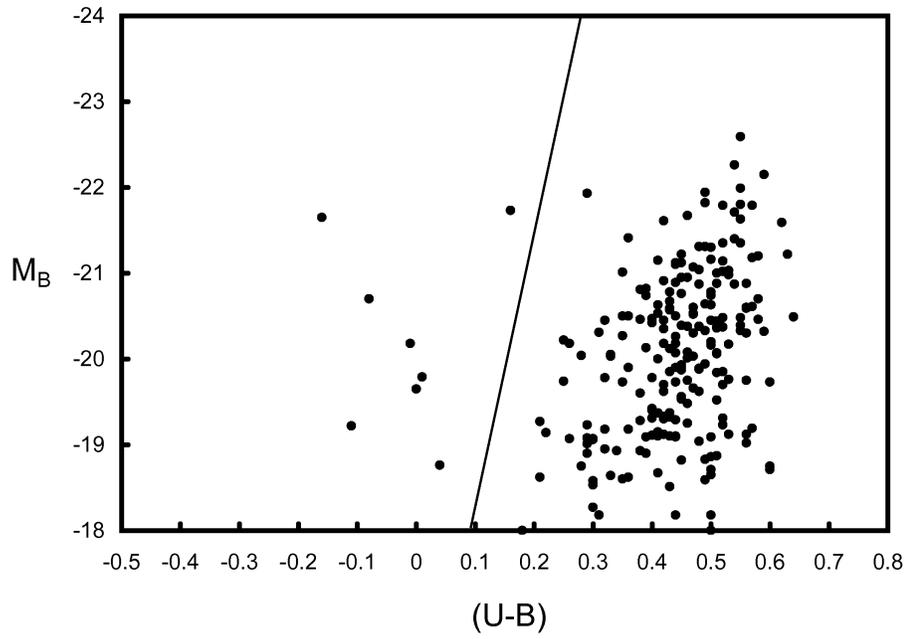}
\caption{Distribution of dustless (D = 0) galaxies shows that most of these objects lie along a broad sequence that is situated $\sim 0.3$ mag to the red of Eqn. (1).}
\end{figure}

\begin{figure}
\plotone{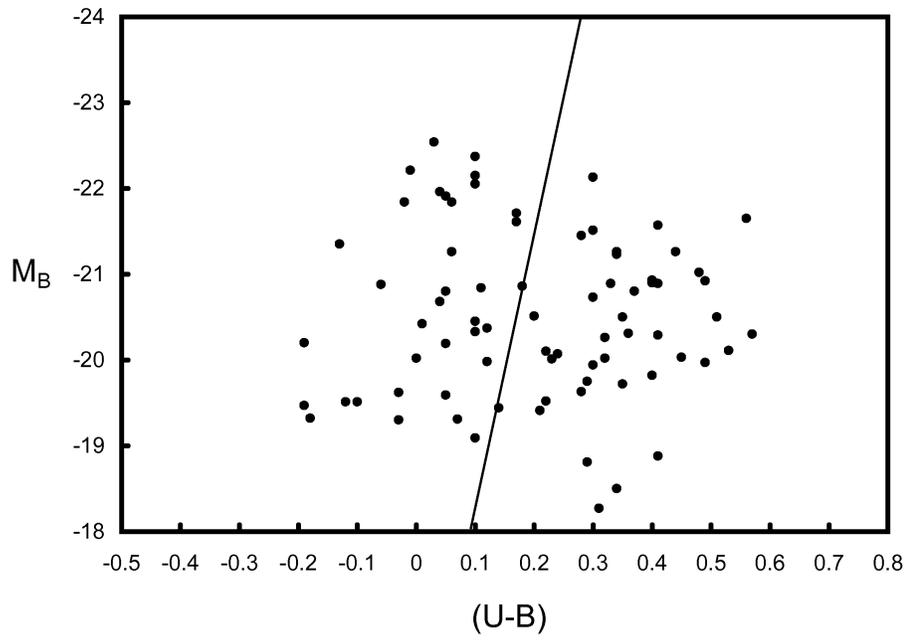}
\caption{Galaxies showing a trace of dust (D = 1) are seen to exhibit a much broader distribution of integrated colors}
\end{figure}

\begin{figure}
\plotone{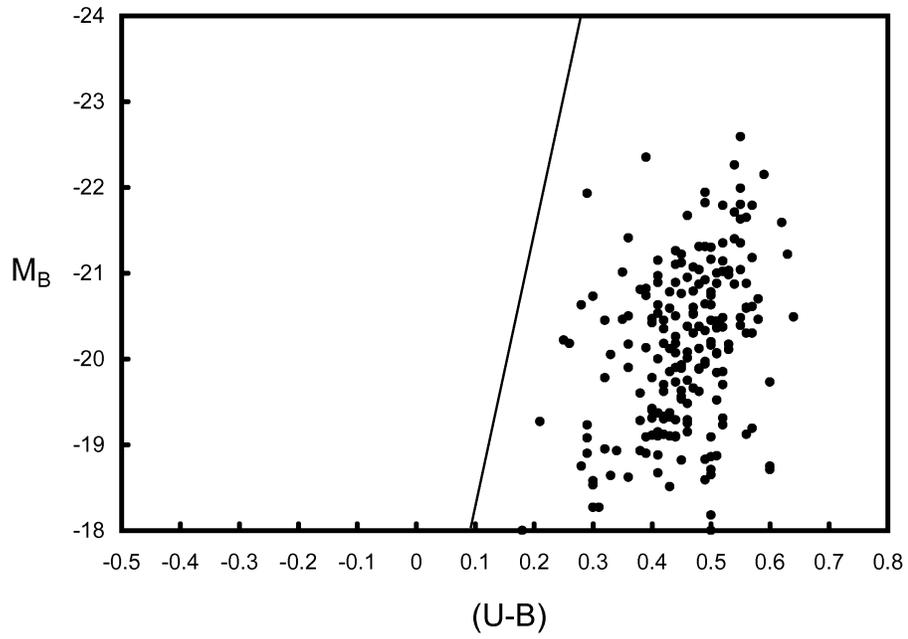}
\caption{Galaxies with no evidence for star formation (S = 0) are mostly located close to a sequence situated $\sim 0.3$ mag to the red of Eqn. (1).}
\end{figure}

\begin{figure}
\plotone{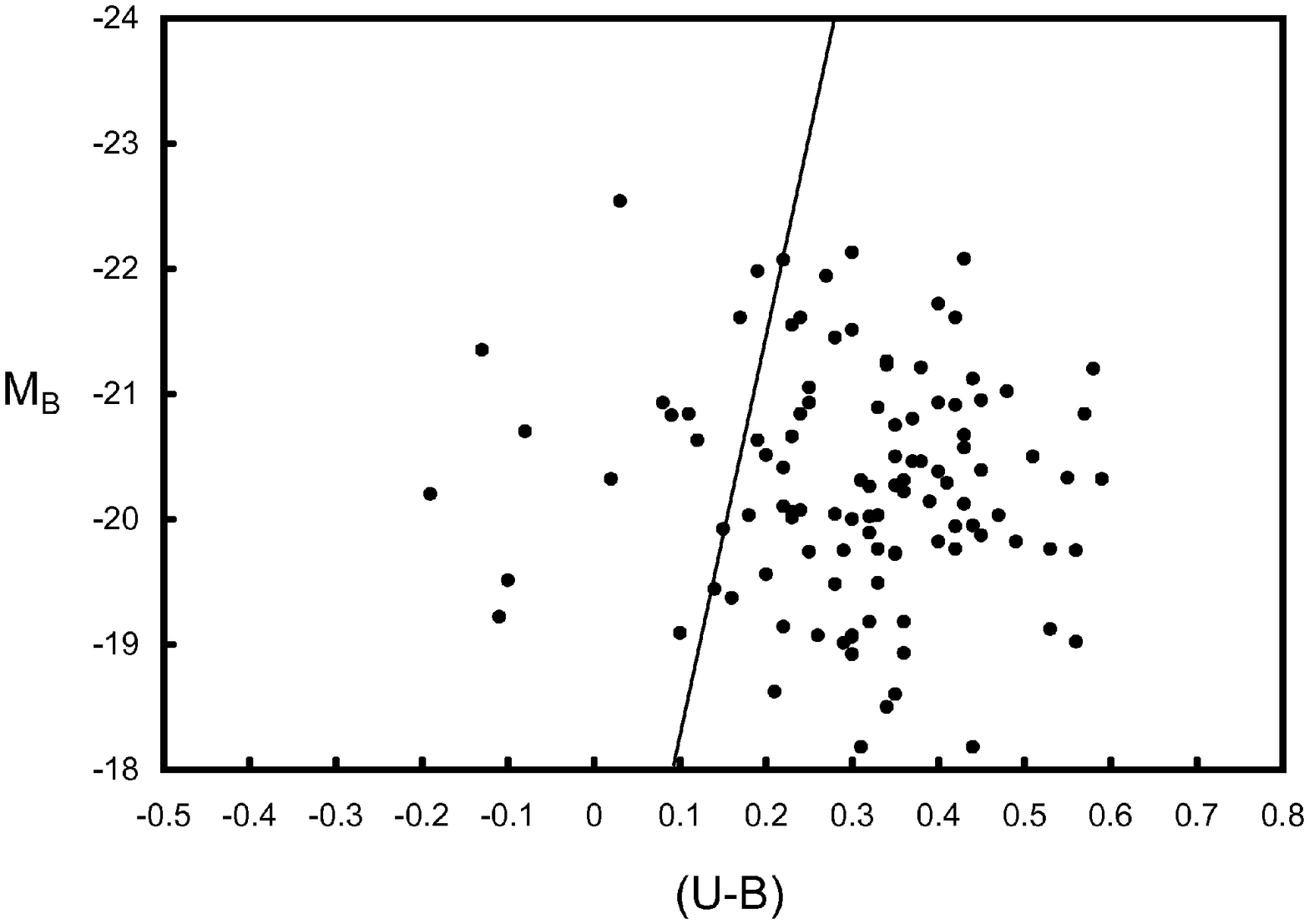}
\caption{Galaxies exhibiting a trace of star formation (S = 1) are typically bluer, and show a much wider dispersion in color, than do those in which no star formation is visible}
\end{figure}


\begin{references}

\reference{}Arnouts, S. et al. 2007, A\&A in press = astro-ph/0705.2436 

\reference{}Baldry, I. K., Balogh, M. L., Bower, R.G., Glazebrook, K., Nichol, R. C., Bamford, S. P. \& Budavari, T. 2006, MNRAS, 373, 469

\reference{}Baldry, I.K., Glazebrook, K., Brinkmann, J., Ivezicd\'{c}, Z. Lupton,R. H., Nichols, R. C., \& Szaley, A. S. 2004, ApJ, 600, 601.

\reference{}Ball, N. M., Loveday, J. \& Brunner, R. J. 2006, MNRAS in press = astro-ph/0610171

\reference{}Cattaneo, A., Dekel, A., Devriendt, J., Guiderdoni, B., \& Blaizot, J. 2006, MNRAS, 370, 1651

\reference{}Chom, Y-Y., Park, C. \& Vogeley, M. S. 2007, ApJ, 658, 884

\reference{}Conselice, C. J. 2006 MNRAS, 373, 1389 

\reference{}de Vaucouleurs, G. 1959, {\it Handbuch der Physik}, 53, 275

\reference{}de Vaucouleurs, G., de Vaucouleurs, A, Corwin, H. G., Buta, R. J., Paturel, G., \& Fouqu\'{e}, P. 1991. {\it Third  Reference Catalog of Bright Galaxies}, Berlin: Springer - Verlag

\reference{}Dressler, A. 1980, ApJ, 236, 351

\reference{}Faber, S. M. et al. 2006 ApJ(submitted) = astro-ph-0506044

\reference{}Hogg, D. W. et al. 2004, ApJ, 601, L29

\reference{}Hubble, E. , 1926, ApJ, 64, 321

\reference{}Hubble, E. , 1936, {\it The Realm of the Nebulae}, New Haven: Yale University Press

\reference{}Kampakoglou, M. \& Silk, J. 2007, MNRAS (in press) = astro-ph-0610607v2

\reference{} Lab$\acute{e}$, I. et al. 2007, ApJ in press = astro-ph/0705.3325

\reference{} Park, C., Chom, Y.-Y., Vogeley, M. S., Gott, R. \& Blanton, M. R. 2007, ApJ, 658, 898


\reference{}Renzini, A. 2006, ARAA, 44, 141

\reference{}Sandage, A. 1961, {\it The Hubble Atlas of Galaxies}, Washington: Carnegie Institution of Washington

\reference{}Sandage, A. 2005, ARAA, 43, 581

\reference{}Sandage, A. \& Bedke, J. , 1994, {\it The Carnegie Atlas of Galaxies, Volumes} I \& II, Washington: Carnegie Institution of Washington

\reference{}Sandage, A., \& Tammann, G. A. 1981, {\it A Revised Shapley-Ames Catalog of Bright Galaxies}, Washington: Carnegie Institution of Washington

\reference{}van den Bergh, S. 1960a, ApJ, 131, 215

\reference{}van den Bergh, S. 1960b, ApJ, 131, 558

\reference{}van den Bergh, S. 1960c, Pub. David Dunlap Obs.,2, 159

\reference{}van den Bergh, S. 1998, {\it Galaxy Morphology and Classification, Cambridge}: Cambridge University Press

\reference{}van den Bergh, S. 2007, Nature 445, 265

\reference{}Wang, Y., Yang, X., Mo, H. J., \& van den Bosch, F. C. 2007, astro-ph/0703253

\reference{}Whiting, A. B., Hau, G. K. T., Irwin, M. \& Verdugo, M. 2007, AJ, 133, 715.

\end{references}
\end{document}